\documentclass[12pt]{iopart}
\usepackage[dvips]{graphicx}

\begin{document}

\title{Cranking mass parameters for fission}
\author{ M. Mirea$^1$ and R.C. Bobulescu$^2$}
\address{$^1$Horia Hulubei National Institute for Physics and Nuclear Engineering, 
Department of Theoretical Physics, P.O. Box MG 6, Bucharest, Romania\\
$^2$University of Bucharest, Faculty of Physics, P.O. Box MG-11, Bucharest,
Romania}
\begin{abstract}
A formalism for semi-adiabatic cranking mass parameters is presented.
For the fission process of $^{234}$U, the time-dependent pairing equations of motion 
were used to calculate
the excitation energy and to extract values of the
cranking inertia. A fission barrier is determined
by minimizing the action trajectory in a five dimensional configuration
space spanned by elongation, necking, deformations of fragments and
mass-asymmetry. The deformation energy is computed 
in the the frame of the microscopic-macroscopic model.
The two center shell model with Woods-Saxon potentials is used in this context.
Values of the inertia for excited fissioning systems are reported.
A dependence between the cranking mass parameters and 
the intrinsic excitation
energy is evidenced.  

\end{abstract}
\maketitle

\section{Introduction}
In microscopic-macroscopic treatments of nuclear fission \cite{nix1}, the whole nuclear system
is characterized by some collective coordinates associated with some
degrees of freedom that determine approximately the behavior
of many intrinsic variables. The basic ingredients in such an analysis
is a shape parametrization that depends on several macroscopic degrees of 
freedom. The generalized coordinates of deformation vary in time
leading to a split of the nuclear system in two separated 
fragments. Thus, these coordinates describe the change in time
of the average field.
Traditionally, in the large amplitude collective motion like fission
or fusion, 
the calculation of the mass parameters are made
within the cranking model \cite{inglis,inglis2}. These parameters are
calculated adiabatically. In reality, the nucleus can be excited
with respect the ground state at any deformation. The excitation energy
is shared between the collective kinetic energy of the fragments and
the intrinsic energy released finally through gamma-ray and neutron emission.
As mentioned in Ref. \cite{schutte}, no unique separation between these two
kind of dumping can be realized. Several attempts were realized in order 
to investigate the change of the cranking mass
parameters as function of the intrinsic excitation energy. In this respect,
investigation based on the cranking model have been done for the
rotational motion \cite{funny} and for the behavior of the temperature
dependent mass parameter \cite{iwamoto1,iwamoto,schneider}. A theory
for non-adiabatic cranking was proposed in Ref. \cite{liran}
where the time-dependent many-body Schrodinger equation is solved
and the time dependence of the collective motion is determined
with the classical Lagrange equation of motion.
It is worth to mention that
in Ref. \cite{chaff} an expression is derived for the collective kinetic
energy containing corrections up to the fourth-order in the
collective velocities. The fourth-corrections to the
cranking approximation at hyperdeformations were found very large.
The authors concluded that the perturbation treatment when applied
to fission is invalidated. 

In this paper, a new treatment for the cranking model is
realized that takes into account the intrinsic excitation energy.
In the next section, a semi-adiabatic cranking approximation is derived
for even-even systems
to investigate the general features of the cranking mass parameters
when the nucleus is internally excited. To make the problem
tractable, it is assumed that the
system deforms slowly in time. Therefore, the matrix elements
of the time-derivative of the wave functions are smalls.
Because these matrix elements are responsible for quasi-particle excitations,
our approximation allows to consider the system in a seniority-zero state during
fission. In other words, the contributions to the inertia
 originating from
 seniority-two configurations are considered negligible.
 In section \ref{sect3}, the mass parameters are calculated.
To obtain the inertia,  a least action trajectory
for the evolution of the nuclear system is determined. 
The single-particle energies for protons and
neutrons were calculated along this path.
Then, the intrinsic excitation
energy of the fissioning system is evaluated 
within the time-dependent
pairing equations of motion 
using different values of the inter-nuclear velocities. 
At each deformation, instantaneous values of
the single-particle densities and of the pairing moment components are
deduced. The semi-adiabatic effective mass is evaluated using these values.
Some behavior concerning the dependence of the cranking mass parameters 
versus the intrinsic excitation
energy are extracted. In the last section, a discussion is made.

\section{Formalism}

The single particle motion of a particle in an average field
is governed by the Schr\"odinger equation that includes
a Hamiltonian with pairing residual interactions.
This Hamiltonian depends on some time-dependent collective parameters
$q(t)=\{q_{\nu}(t)\}$ ($\nu=1,...n$), such as the internuclear distances between 
the nascent fragments, the mass-asymmetry, the fragment deformations or the necking parameter:
\begin{equation}
H(t)=\sum_{k>0} \epsilon_{k}[q(t)](a_{k}^{+}a_{k}+a_{\bar k}^{+}a_{\bar k})-G\sum_{k,i>0}a_{k}^{+}a_{\bar{k}}^{+}
a_{i}a_{\bar{i}}.
\label{ham1}
\end{equation}
where $\epsilon_{k}$ are single-particle energies,
$G$ is an monopole pairing interaction constant and $a^+_k$ denote creation
operators.
In order to obtain the time dependent pairing
equations of motion, we shall start from the variational
principle taking the following energy functional
\begin{eqnarray}
{\cal{L}}=\langle\varphi \mid H-i\hbar{\partial\over \partial t}-
\lambda \hat{N}\mid \varphi\rangle
\label{expr0}
\end{eqnarray}
and by assuming the many-body state formally expanded as a superposition of
time dependent BCS seniority-zero and seniority-two adiabatic wave functions
\begin{eqnarray}
\mid\varphi(t)\rangle =c_{0}(t)\mid\phi_{\rm BCS}\rangle
 +\sum_{j,l}c_{jl}(t)\alpha_{j}^{+}\alpha_{\bar{l}}^{+}\mid\phi_{\rm BCS}\rangle
\label{wf2}
\end{eqnarray}
where 
\begin{equation}
\mid \phi_{\rm BCS}\rangle=
\prod_{k}(u_{k}+v_{k}a_{k}^+a_{\bar k}^+)\mid 0\rangle
\label{wf3}
\end{equation}
is the seniority-zero Bogoliubov wave function.
The excited seniority-two configurations are obtained by mean of 
quasiparticle creation and annihilation operators: 
\begin{eqnarray}
\alpha_{k}=u_{k}a_{k}-v_{k}a_{\bar k}^{+};&
~~~\alpha_{\bar{k}}=u_{k}a_{\bar k}+v_{k}a_{ k}^{+};\\
\alpha_{k}^{+}=u_{k}a_{k}^{+}-v_{k}^{*}a_{\bar k}; &
~~~\alpha_{\bar{k}}^{+}=u_{k}a_{\bar k}^{+}+v_{k}^{*}a_{ k};\nonumber
\label{anho2}
\end{eqnarray}
In definitions (\ref{expr0}) and (\ref{wf2}), $c_{0}$ and $c_{jl}$ are amplitudes of the
two kinds of configurations, $\lambda$ is the chemical potential, and 
$\hat{N}$ is the particle number operator. Because only the relative phase
between the parameters $u_{k}$ (vacancy amplitudes) and $v_{k}$ 
(occupation amplitudes) matters, in the following
 $u_{k}$ is considered to be a real quantity and $v_k$ a complex one. 
To minimize the functional, the
expression (\ref{expr0}) is derived with respect the independent
variables $v_{k}$, $v_{k}$, 
$c_{0}$,  $c_{jl}$ (the amplitudes of the wave function),  
together with their complex conjugates,
 and the resulting equations
are set to zero.

As deduced in Refs. \cite{nix,franta} and as detailed in the Appendix,
for a seniority-zero nuclear system, the
pairing equations of motion are:
\begin{eqnarray}
i\hbar \dot{\rho}_{k}&=\kappa_{k}\Delta_{0}^{*}-
\kappa_{k}^{*}\Delta_{0};\label{hfb}\nonumber\\
i\hbar \dot{\kappa}_{k}&=\left(2\rho_{k}-1\right)\Delta_{0}+
2\kappa_{k}\left(\epsilon_{k}-\lambda\right)
   -2G\rho_{k}\kappa_{k};
\label{tdp}
\end{eqnarray}
This system is sometimes called time dependent 
Hartree-Fock-Bogoliubov equations \cite{franta}. These equations
were already used to determine the intrinsic excitation energy 
in fission \cite{mireanp,mirealz} or to investigate the pair-breaking
mechanism \cite{mireapl}.
The following notations are used in Eqs. (\ref{hfb}) and in 
the remaining part of the article:
\begin{eqnarray}
\Delta_{0}=G\sum_{k}\kappa_{k};& \Delta_{jl}=G\sum_{k\ne j,l}\kappa_{k}; \nonumber\\
\kappa_{k}=u_{k}v_{k}; &  
\rho_{k}=\mid v_{k}\mid^{2};\label{notatii}
\end{eqnarray}
where $\Delta_0$ is the gap parameter for the seniority-zero
state while $\Delta_{jl}$ are related to seniority-two states.
$\kappa_{k}$ are pairing moment components and $\rho_{k}$ are
single particle densities.

An estimate
of the intrinsic seniority-zero state excitation energy \cite{nix} 
can be obtained with the relation
\begin{equation}
E^*=E_{0}-E_{\rm BCS}
\label{exc}
\end{equation}
where
\begin{equation}
E_{0}=\langle \phi_{\rm BCS}\mid H-\lambda\hat N\mid\phi_{\rm BCS}\rangle
 =2\sum_{k} \rho_{k}(\epsilon_{k}-\lambda)
-{\mid \Delta_{0}\mid^{2}\over G}-G\sum_{k}\rho_{k}^{2};
\label{uiy}
\end{equation}
is the expected value of the Hamiltonian (\ref{ham1}) for
the seniority-zero state and $E_{\rm BCS}$ is the stationary
energy obtained by replacing $\kappa_{k}$ and $\rho_{k}$ within
the time-independent BCS parameters $\tilde\kappa_k$ and 
$\tilde\rho_k$ in formula (\ref{uiy}).

As indicated in the Appendix, the next
semi-adiabatic cranking formula can be obtained
for the effective mass parameters $B$:

\begin{eqnarray}
\label{crr}
\fl B_{\nu\mu}=B_{1\nu\mu}+B_{2\nu\mu}\nonumber\\
\fl =2\hbar^2\sum_{m,n\ne m}
{ (E_{mn}-E_0)\mid {\kappa_m\sqrt{\rho_{m}}\mid\kappa_{n}\mid
\over\mid\kappa_{m}\mid\sqrt{\rho_{n}}}-{\kappa_n\sqrt{\rho_{n}}\mid\kappa_{m}\mid
\over\mid\kappa_{n}\mid\sqrt{\rho_{m}}
}\mid^2
\langle m\mid{\partial H\over\partial q_{\nu}}\mid n\rangle
\langle n\mid{\partial H\over\partial q_{\mu}}\mid m\rangle
 \over (E_{mn}-\sum_{k\ne m,n}T_k-E_{0}+\sum_{k}T_{k})^2
(\epsilon_{m}-\epsilon_{n})^2}\\
\fl +2\hbar^{2}\sum_{m}
{ (E_{mm}-E_0)
({\kappa_m\over\rho_m}{\partial\rho_m\over\partial q_\nu}-
{\kappa_m\over\kappa_m^*}{\partial\kappa_m^*\over\partial q_\nu})
({\kappa_m^*\over\rho_m}{\partial\rho_m\over\partial q_\mu}-
{\kappa_m^*\over\kappa_m}{\partial\kappa_m\over\partial q_\mu})
\over (E_{mm}-\sum_{k\ne m}T_k+T_m-E_{0}+\sum_{k}T_{k})^2}\nonumber
\end{eqnarray}
where
the values of $\rho_{k}$ and $\kappa_{k}$ are solutions of the 
time dependent pairing equations (\ref{tdp})i and the index $\nu$
is associated to the collective coordinate $q_\nu$.. 
$E_{jl}$ are exactly
the expected values of the Hamiltonian (\ref{ham1}) for
 seniority-two 
configurations:
\begin{eqnarray}
E_{jl}&=\langle\alpha_{j}^+\alpha_{\bar l}^+ 
\phi_{\rm BCS}\mid H-\lambda\hat N\mid
\alpha_j^+\alpha_{\bar l}^+\phi_{\rm BCS}\rangle\label{euri1}\\
& =2\sum_{k\ne j,l} \rho_{k}(\epsilon_{k}-\lambda)
-{\mid \Delta_{jl}\mid^{2}\over G}
  -G\sum_{k\ne j,l}\rho_{k}^{2}+
\mid\epsilon_{j}-\lambda\mid+\mid\epsilon_{l}-\lambda\mid;\nonumber\\
\label{euri}
E_{jj}&=\langle\alpha_{j}^+\alpha_{\bar j}^+ 
\phi_{\rm BCS}\mid H-\lambda\hat N\mid
\alpha_j^+\alpha_{\bar j}^+\phi_{\rm BCS}\rangle
=2\sum_{k\ne j}\rho_{k}(\epsilon_{k}-\lambda)\\
&+
2u^2\mid\epsilon_{j}-\lambda\mid
-{\mid \Delta_{j}\mid^{2}\over G}
  +\kappa_{j}\Delta_{jj}+\kappa_{j}^{*}\Delta_{jj}^{*}
-G\sum_{k\ne j}\rho_{k}^{2}+\mid\kappa_{j}\mid^4/\rho_{j}^2;\nonumber
\end{eqnarray}
and $T_{k}$  are energy terms associated to single-particle states:
\begin{equation}
T_{k}
=2\rho_{k}(\epsilon_{k}-\lambda)
-2G\rho_{k}^{2}
 +{\kappa_{k}\Delta_{0}^{*}
+\kappa_{k}^{*}\Delta_{0}\over 2}
\left({\rho_{k}^{2}\over \mid\kappa_{k}\mid^{2}}-1
\right);
\label{tk}
\end{equation}

The mass parameters can be obtained from formula (\ref{crr}) 
only if the derivatives $\partial \rho_m/\partial q_\nu$
and $\partial \kappa_m/\partial q_\nu$ are supplied.
In the BCS stationary state, the derivatives 
$\partial \tilde\rho_m/\partial q_\nu$
and $\partial \tilde\kappa_m/\partial q_\nu$
depend only on the collective variables $q_\nu$
and the cranking formula can be easily obtained.
On the other hand, if the system follows a trajectory 
in the configuration space and the motion is characterized
by some given 
 collective velocities, then the derivatives depend
also on $\dot q_\nu$. Therefore,
the second term in Eq. (\ref{crr}) depends on the history
of the system and can be calculated only if
 the variations of $\rho_k$ and $\kappa_{k}$ are
known. The
inertia along the trajectory is
\begin{eqnarray}
\label{fin}
 B=&\sum_{\nu,\mu}B_{\nu\mu}{\partial q_\nu\over\partial R}
 {\partial q_\mu\over\partial R}\\
\fl &=\sum_{\nu,\mu}B_{1\nu\mu}{\partial q_{\nu}\over\partial R}
{\partial q_{\mu}\over\partial R}+
2\hbar^2\sum_{m}{ (E_{mm}-E_0)
\mid{\kappa_m\over\rho_m}{\partial\rho_m\over\partial R}-
{\kappa_m\over\kappa_m^*}{\partial\kappa_m^*\over\partial R}\mid^2
\over (E_{mm}-\sum_{k\ne m}T_k+T_m-E_{0}+\sum_{k}T_{k})^2}\nonumber
\end{eqnarray}
where one coordinate $R$ is taken as the independent variable.
The other coordinates $q_\nu$ are taken to be function of $R$ \cite{funny}.
From definition, the collective kinetic energy is $E_{c}= B \dot R^2/2$ and
the last term in Eq. (\ref{fin}) becomes:
\begin{equation}
E_{c0}=\hbar^2\sum_{m}{ (E_{mm}-E_0)
\mid{\kappa_m\over\rho_m}\dot\rho_m-
{\kappa_m\over\kappa_m^*}\dot\kappa_m^*\mid^2
\over (E_{mm}-\sum_{k\ne m}T_k+T_m-E_{0}+\sum_{k}T_{k})^2}
\label{sec}
\end{equation}
This term depends only on the derivatives with respect to time
$\dot\kappa_m$ and $\dot\rho_m$. Their expressions are given
by the coupled channel system of equations
 (\ref{tdp}). That means, the term (\ref{sec}) is 
practically independent on the collective velocity and depends
only on the values of $\epsilon_m$,
 $\kappa_{m}$ and $\rho_m$, that is on the
structure of the system and its intrinsic excitation
and represents the minimal
collective kinetic energy. Therefore, the term (\ref{sec})
can be viewed as an ground
collective kinetic energy. In other words, if the nucleus is internally
excited and the derivatives of the probabilities $\dot\kappa$, $\dot\rho$
are different
from zero, the system possesses a minimal collective kinetic energy.
The term (\ref{sec}) can be considered as responsible for energy transfer between
intrinsic and collective degrees of freedom. 

\section{Results}

\label{sect3}

\begin{figure}
\resizebox{0.5\textwidth}{!}{\includegraphics{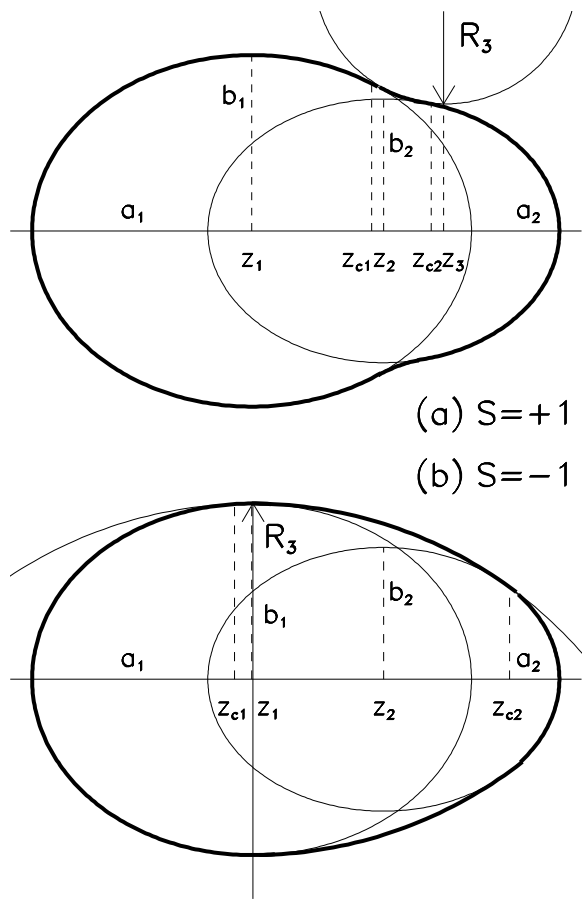}}
\caption{Nuclear shape parametrization.}
\label{fig1}
\end{figure}

The calculation addresses the fission of $^{234}$U.
As already mentioned, the basic ingredient in our analysis 
is the nuclear
shape parametrization.
The nuclear
shape parametrization used in the following
is given by two ellipsoids of different sizes smoothly joined by a third
surface obtained \cite{mirea2}
by the rotation of a circle around the axis of
symmetry. Five degrees of freedom characterize this parametrization:
the elongation given by the inter-nuclear distance $R$ between
the centers of the ellipsoids, the two deformations
of the nascent fragments characterized by their eccentricities 
$\varepsilon_{i}=\sqrt{1-b_{i}^2/a_{i}^2}$ $(i=1,2)$,
 the mass asymmetry given by the ratio between the
major semi-axis of the fragments $\eta=a_{1}/a_{2}$, 
and the necking parameter related to the median curvature
$C=s/R_{3}$ ($R_{3}$ being the radius of the intermediate circle and $S$ the
sign associated to the curvature). The meaning of all geometric parameters
can be understand by inspecting the Fig. \ref{fig1}.

\begin{figure}
\resizebox{0.6\textwidth}{!}{\includegraphics{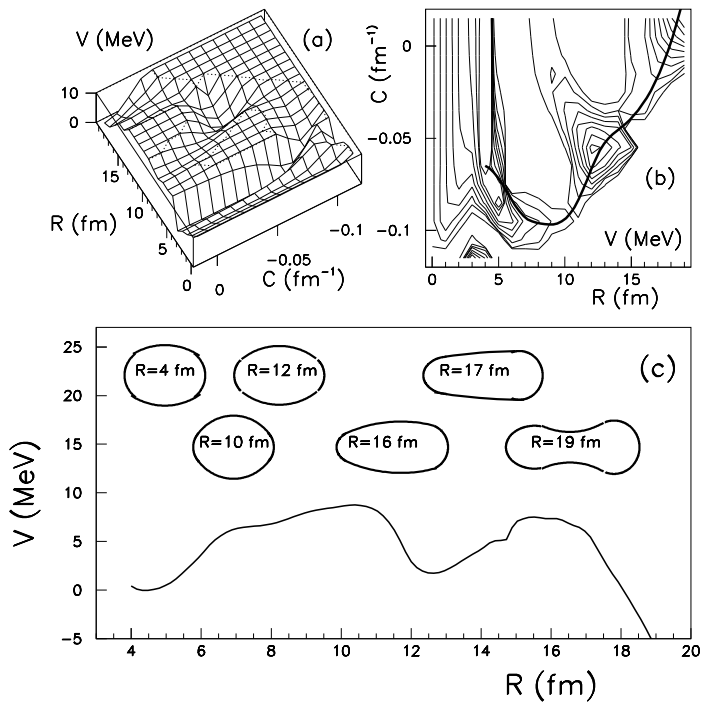}}
\caption{(a) Minimal values of the deformation energy
in MeV as function of the necking coordinate $C$ and 
the elongation $R$ for $^{234}$U. (a) Contours of the
deformation energy in step of 1 MeV. The least action trajectory
is superimposed. (c) Potential barrier. Some shapes
obtained during the fission process together with
the values of the elongation $R$ are inserted.
}
\label{fig2}
\end{figure}

As specified in Ref. \cite{funny}, first of all, a calculation of
the fission trajectory
in our five-dimensional configuration space, beginning with the ground-state of the system up to the exit point of the barrier must be performed.
This can be done
by minimizing the action integral. For this purpose, two ingredients
are required: the deformation energy $V$ and the tensor of the effective mass.
The deformation energy was obtained \cite{nix1}  by summing the liquid drop
energy with the shell and the pairing corrections.
The macroscopic energy is obtained
in the framework of the Yukawa plus exponential model \cite{davies}
extended for binary systems with different charge densities
\cite{ejpa}. The Strutinsky microscopic corrections
 were computed on the basis of the Woods-Saxon
superasymmetric two-center
shell model \cite{mirea2}. This model gives the
single particle level diagrams by diagonalizing a Woods-Saxon potential,
corrected within spin-orbit and Coulomb terms, in the analytic eigenvalue basis
of the two center semi-symmetric harmonic model \cite{mnpa,gr}.
The effective mass is computed within the cranking adiabatic 
approximation as given in \cite{funny}.
After minimization,
the dependences between the generalized coordinates 
$q_{\nu}$ ($\nu=1,...5$)
in the region comprised between the parent ground state configuration and the
exit point of the external fission barrier
supply the least action trajectory.
The ground-state corresponds to the lowest deformation
energy in the first well.
The least action trajectory is obtained
within
a numerical method.
Details about the numerical procedure of minimization
and about the model can be found
in Refs. \cite{mirealz,mirea2,mirea3} and references therein.
 Plots of the minimal deformation energy surface as function of
 the necking coordinate $C$ and the elongation $R$ are
displayed in Figs. \ref{fig2} (a) and (b). In Fig. \ref{fig2} (b)
the minimal action trajectory is also plotted.
The resulting $^{234}$U fission barrier is plotted on
Fig. \ref{fig2} (c) as function of the distance between the
centers of the nascent fragments $R$. Some nuclear shapes
obtained along the minimal action trajectory are inserted in
the plot. 

\begin{figure}
\resizebox{0.6\textwidth}{!}{\includegraphics{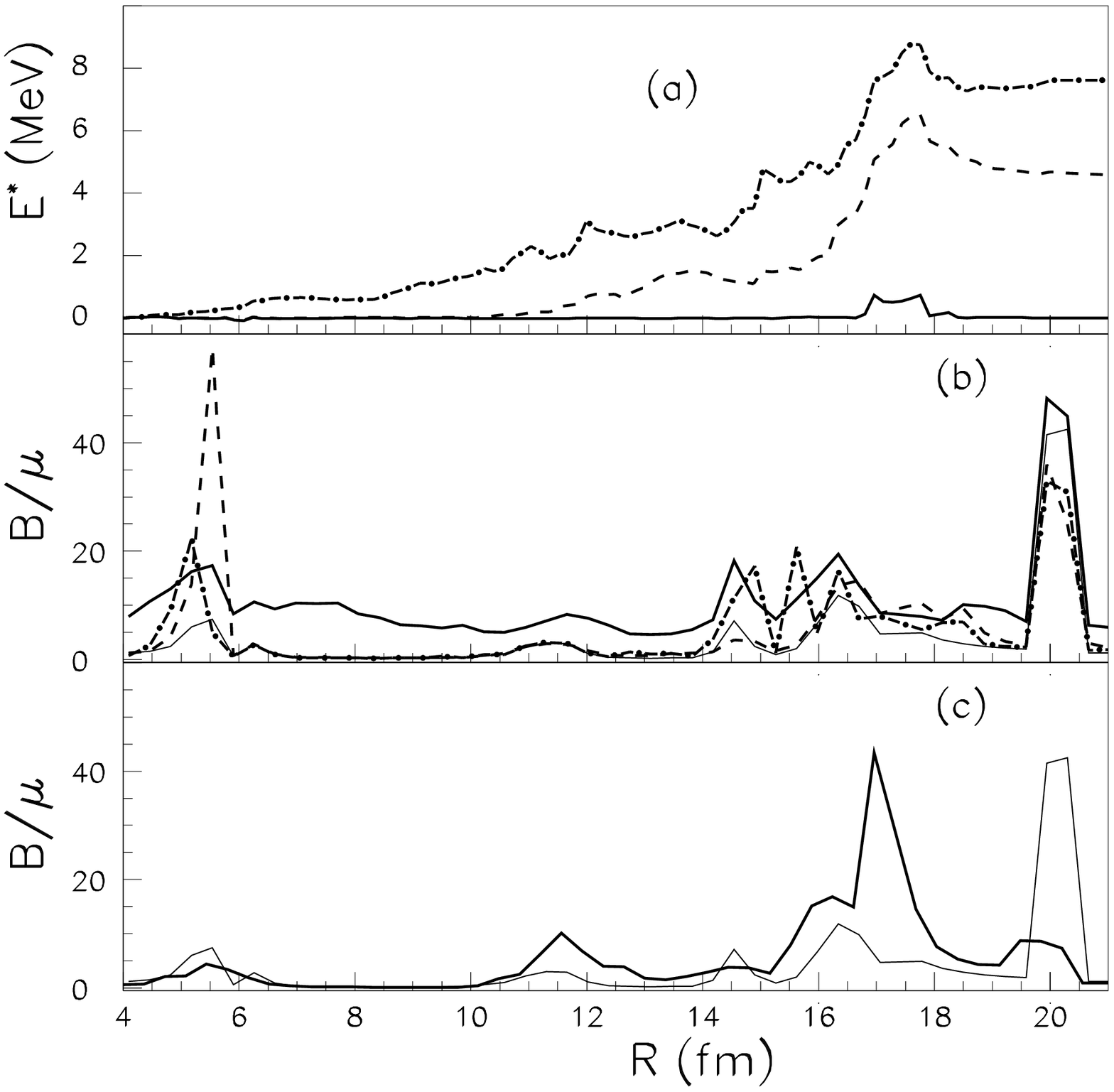}}
\caption{(a) Intrinsic excitation energies along the minimal
action trajectory as function
of the elongation $R$ for three values of the internuclear
distance velocity $\dot R$: full line $\dot R=1\times10^{4}$ fm/fs,
 dashed line $\dot R=1\times10^{5}$ fm/fs and dot-dashed line
$\dot R=1\times10^{6}$ fm/fs. (b) The thick lines represent
the inertia $B$ divided by the reduced mass $\mu$
for the three velocities taken into consideration
as function of $R$.
The same line types are used as in panel (a) for the values
of the velocities. The thin line
corresponds to the inertia computed within BCS parameters
$\tilde\kappa$ and $\tilde\rho$. (c) Comparison
between the inertia calculated within formula (\ref{fin}) with
stationary values $\tilde\kappa$ and $\tilde\rho$
(thin line) and the classical 
cranking formula \cite{funny} (thick line).
Asymptotically, the two inertia reaches approximately the
reduced mass $\mu$.
}
\label{fig3}
\end{figure}

The realistic  level scheme along the least action
trajectory were also obtained within the superasymmetric Woods-Saxon
two-center shell model. Within the energy diagrams, the system
(\ref{tdp}) is solved having as initial conditions the stationary
BCS state in the ground state configuration. 
Three values of the internuclear velocity $\dot R$ were 
taken into account: $1\times 10^{4}$, $1\times 10^{5}$
and $1\times 10^{6}$ fm/fs, corresponding to a time
to penetrate the barrier comprised between $[10^{-18},10^{-20}]$ s.
In Fig. \ref{fig3} (a), the intrinsic excitation energy given
by Rel. (\ref{exc}) is plotted as
function of the internuclear distance for the three values of $\dot R$.
If the internuclear velocity increases, the excitation energy becomes
larger.

The inertia along the trajectory was calculated
within three different approaches: the adiabatic cranking
model of Ref. \cite{funny,leder,mbm}, the formula (\ref{fin}) within
stationary BCS parameters $\tilde \kappa$ and $\tilde \rho$ 
 and the same formula within $\kappa$ and $\rho$ values given
by Eq. (\ref{tdp}).
The total effective mass is the sum of partial values obtained 
for neutron and proton subsystems.

In Fig. \ref{fig3} (c), the inertia along the minimal action path
calculated within the adiabatic cranking model is displayed
with a thick line. In the ground state of the first well, the inertia
is very small. Large values of of the cranking inertia
are obtained around the exit point of the outer fission barrier,
close to the scission configuration. A value approaching the reduced mass
is obtained after the scission point. A similar behavior, 
exhibiting very large values of the inertia around the touching
configuration, was obtained for light systems in Ref. \cite{torres},
where a version of the Woods-Saxon two center shell model based
on the molecular orbital approach is used. In the same plot,
a thin line gives the inertia obtained within formula (\ref{fin})
by using the stationary values $\tilde\kappa$ and $\tilde\rho$.
This semi-adiabatic inertia exhibits a similar shell structure
as the adiabatic cranking model. The semi-adiabatic model gives in general
lower values than the adiabatic cranking one, excepting the regions
of the ground state and of the scission point where the values are
larger.

In Fig. \ref{fig3} (b) a comparison is made between inertia computed
within relation (\ref{fin}) for different values of the internuclear
velocity. The same values of the internuclear velocities $\dot R$
were taken as those used to determine the excitation energies. The
inertia along the minimal action trajectory has the largest values
for the lowest $\dot R$ taken into consideration. In general, the
values obtained for $\dot R=1\times 10^4$ fm/fs represent an upper limit
for the magnitude of the inertia. It must be mentioned that for this
collective velocity, the calculated intrinsic excitation energy
is practically negligible. For higher values of $\dot R$, when the
excitation energy becomes important, the magnitude of the inertia
is considerably lower. The values obtained for 
$\dot R=1\times 10^5$ fm/fs and $\dot R=1\times 10^6$ fm/fs
are situated in the vicinity of those obtained within the
stationary values. For the three collective velocities, the shell
structure resemble.

\section{Discussion}
                              
In Ref. \cite{schneider}, a temperature dependent cranking model 
that includes a parameter
for the dissipation was used. It was shown that the increase of the
temperature smoothen and lower the mass parameter. In the high temperature
limit, the cranking results are quite close to the irrotational
flow results. Our calculations also show that the inertia decreases
when the excitation energy increases.

Several approximations were derived in the literature to
compute the collective inertia. In the frame of the response theory
 \cite{hoff} the Hamiltonian is expanded around a particular value
of the macroscopic coordinate and the first derivative is
treated with the time dependent perturbation approach.
Another method is related to the generator coordinate method
proposed in Ref. \cite{brink} within the Gaussian Overlap
Approximation (GOA). This method allows to obtain a representation of
any operator in the collective space. In general, the GOA mass 
is about 2/3 times smaller than the adiabatic cranking mass \cite{gozdz}
but both quantities exhibits a similar shell structure.
Adiabatic mass parameters for fission were derived also from 
the Time-Depedent Hartree-Fock-Bogololiubov (TDHFB) theory \cite{baran}.
A comparison between values of the collective mass tensor
obtained  with three diferent models, i.e.,
cranking, GOA and adiabatic TDHFB, showed that the 
adiabatic TDHFB mass exhibits more pronounced variations than
the cranking and the GOA masses. As expected, the GOA gives smaller
values than the cranking model. Our results show that for stationary values
of $\tilde\kappa$ and $\tilde\rho$,  the inertia is lower than 
the adiabatic cranking values.

In Ref. \cite{dudek} the second $0^+$ collective energy level
was calculated for the rare-earth and actinide nuclei using
the Bohr-Sommerfeld quantization rule. The results overestimates
the experimental values in average within a factor 2. This 
discrepancy ca be caused either by the shape of the potential
in the ground state or either by a systematic too low value
of the cranking inertia. In our calculations, 
in the first well has larger values
than that obtained within the adiabatic cranking model.

In conclusion, a semi-adiabatic formalism based
on the time-dependent pairing equations was described.
The values of the inertia obtained within this model
exhibit a strong dependence on the intrinsic excitation
energy. In general, the semi-adiabatic inertia have a similar
shell structure as the adiabatic cranking model.
\section{Acknowledgments}

Work supported by the Grant CNCSIS IDEI 512 of the Romanian Ministry
of Education and Research.

\section{APPENDIX}

In order to simplify the calculations, the variations of the 
parameter $u_{k}$ and $v_{k}$ as function of the seniority 
number, called blocking effect,
is neglected. A treatment involving the blocking effect was
realized in Ref. \cite{mirea2} in order 
to generalize the Landau-Zener effect and in Ref. \cite{mireapl}
to evidence a new dynamical pair breaking effect.
Within the wave function (\ref{wf2}), using the 
notation (\ref{wf3}) and the following identities
\begin{equation}
\langle c_{0}\phi_{\rm BCS}\mid {\partial\over\partial t}\mid c_{0}\phi_{\rm BCS}\rangle
=c_{0}^{*}\dot{c_{0}}+
\mid c_{0}\mid^{2}\sum_{k}(u_{k}\dot u_{k}+v_{k}^{*}\dot v_{k});
\end{equation}
                    
\begin{eqnarray}
\fl\langle\sum_{j,i\ne j}c_{ji}\alpha_{j}^{+}\alpha_{\bar i}^{+}\phi_{\rm BCS}
\mid {\partial \over\partial t}\mid 
\sum_{k,l\ne k}c_{kl}\alpha_{k}^{+}\alpha_{\bar l}^{+}\phi_{\rm BCS}\rangle
 = \sum_{j,l\ne j}[
c_{jl}^{*}\dot{c_{jl}}+
\mid c_{jl}\mid^{2}\sum_{k\ne j,l}(u_{k}\dot u_{k}+v_{k}^{*}\dot v_{k})]\nonumber\\
 + \sum_{j,i\ne j}\sum_{k,l\ne k}c_{ji}^*c_{kl}[(u_{k}u_{j}+v_{k}^{*}v_{j})
\langle j\mid {\partial\over\partial t}\mid k\rangle\delta_{il}\\
     +
(u_{i}u_{l}+v_{l}^{*}v_{i})
\langle\bar i\mid {\partial\over\partial t}\mid \bar l\rangle\delta_{jk}];\nonumber
\end{eqnarray}
                        
\begin{equation}
\langle c_{jl}\alpha_{j}^{+}\alpha_{\bar l}^{+}\phi_{\rm BCS}
\mid {\partial \over\partial t}\mid 
c_{0}\phi_{\rm BCS}\rangle=
c_{jl}^{*}c_{0}(u_{j}v_{l}-v_{j}u_{l})
\langle j\mid {\partial\over\partial t}\mid l\rangle;
\end{equation}

\begin{equation}
\langle c_{0}\phi_{\rm BCS}\mid {\partial \over\partial t}\mid 
c_{jl}\alpha_{j}^{+}\alpha_{\bar l}^{+}\phi_{\rm BCS}\rangle=
-c_{0}^{*}c_{jl}(u_{j}v_{l}^{*}-v_{j}^{*}u_{l})
\langle j\mid {\partial\over\partial t}\mid l\rangle;
\end{equation}

\begin{equation}
\langle c_{jl}\alpha_{j}^{+}\alpha_{j}^{+}\phi_{\rm BCS}\mid {\partial\over\partial t}\mid 
c_{0}\phi_{\rm BCS}\rangle=
-c_{jl}^{*}c_{0}(u_{j}\dot v_{j}-v_{j}\dot u_{j});
\end{equation}
\begin{equation}
\langle c_{0}\phi_{\rm BCS}\mid {\partial\over\partial t}\mid 
c_{jj}\alpha_{j}^{+}\alpha_{j}^{+}\phi_{\rm BCS}\rangle=
-c_{0}^{*}c_{jj}(u_{j}\dot v_{j}^{*}-v_{j}^{*}\dot u_{j});
\end{equation}
the energy functional (\ref{expr0}) becomes
\begin{eqnarray}
\fl{\cal{L}}=\mid c_{0}\mid^{2}\left\{\sum_{k}2\mid v_{k}\mid^2(\epsilon_{k}-\lambda)-
G\mid\sum_{k}u_{k}v_{k}\mid^2-G\sum_{k}\mid v_{k}\mid^4\right\}\nonumber\\
+\sum_{j,l\ne j}\mid c_{jl}\mid^{2}\left\{\sum_{k\ne j,l}2\mid v_{k}\mid^2
(\epsilon_{k}-\lambda)+\mid\epsilon_{j}-\lambda\mid+
\mid\epsilon_{k}-\lambda\mid
\nonumber\right.\\
-\left.
G\mid\sum_{k\ne j,l}u_{k}v_{k}\mid^2-G\sum_{k\ne j,l}\mid v_{k}\mid^4\right\}\nonumber\\
+\sum_{j}\mid c_{jj}\mid^{2}\left\{\sum_{k\ne j}2\mid v_{k}\mid^2
(\epsilon_{k}-\lambda)+2u^2\mid\epsilon-\lambda_{jj}\mid\nonumber\right.\\
-\left.
G\mid\sum_{k\ne j}u_{k}v_{k}\mid^2+u_jv_jG\sum_{k\ne j}(u_{k}v_{k})+
u_jv_j^*G\sum_{k\ne j}(u_{k}v_{k}^*)
-G\sum_{k\ne j}\mid v_{k}\mid^4-u_{j}^{4}\right\}\nonumber\\
-i\hbar\left\{c_{0}^*\dot c_{0}+\mid c_{0}\mid^{2}\sum_{k}{1\over 2}
(v_{k}^*\dot v_{k}-\dot v_{k}^*v_{k})\right.\nonumber\\
+\sum_{j,l\ne j}\left[c_{jl}^*\dot c_{jl}+\mid c_{jl}\mid^{2}\sum_{k\ne j,l}{1\over 2}
(v_{k}^*\dot v_{k}-\dot v_{k}^*v_{k})\nonumber\right]\\
+\sum_{j}\left[c_{jj}^*\dot c_{jj}+\mid c_{jj}\mid^{2}\left[\sum_{k\ne j}{1\over 2}
(v_{k}^*\dot v_{k}-\dot v_{k}^*v_{k})-
{1\over2}(v_{j}^*\dot v_{j}-\dot v_{j}^*v_{j})\right]\nonumber\right]\\
+\sum_{j,l\ne j}\left[c_{jl}^*c_{0}(u_{j}v_{l}-v_{j}u_{l})+
c_{jl}c_{0}^*(v_{j}^*u_{l}-v_{l}^*u_{j})\right]
\langle j\mid{\partial\over\partial t}\mid l\rangle\nonumber\\
+\sum_{j}\left[c_{jj}^*c_{0}(-u_{j}\dot v_{j}+v_{j}\dot u_{j})+
c_{jj}c_{0}^*(-u_{j}\dot v_{j}^*+v_{j}^*\dot u_{j})\right]\nonumber\\
 +\sum_{j,l\ne j}\sum_{m,n\ne m}c_{jl}^*c_{mn}
\left[(u_{m}u_{j}+v_{m}^*v_{j})
\langle j\mid{\partial\over\partial t}\mid m\rangle\delta_{ln}\right.\nonumber\\
\left.\left. +
(u_{n}u_{l}+v_{n}^*v_{l})
\langle\bar l\mid{\partial\over\partial t}\mid \bar n\rangle
\delta_{jm}\right]\right\},
\label{lagra}
\end{eqnarray}
where the dot represents the time derivative.
In order to obtain the equations of motion, the previous expression must be derived
with respect the independent variables ant their complex conjugates
and the resulting expressions must be set to zero.
The assumption of a slowly 
varying deformation with respect the time will be used. 
In these circumstances, the probability to find the system in a 
real or virtual seniority-two state is much lower than the probability
to find the system in a seniority-zero one. These statements can 
be translated in the following conditions: $\mid c_{ij}\mid^2<<1$ and $\mid c_{0}\mid^2\approx 1$.
In this context, as prescribed in
Ref. \cite{mirea2} and using the derivatives with 
respect $v_{l}$ and $v_{l}^{*}$, two equations are obtained: 
\begin{equation}
\fl -i\hbar\dot v^*=2v_{m}^*(\epsilon_{m}-\lambda)-G
\left[\sum\kappa_{k}\left(-{v_l^*v_l^*\over 2u_l}\right)+
\left(u_{l}-{\rho_{l}\over 2u_{l}}\right)\sum\kappa_k^*+2\rho_lv_l^*\right],
\label{eqd}
\end{equation}
and another for its complex conjugate. The notations (\ref{notatii}) are used.
 From Eq. (\ref{eqd}) and its complex conjugate, the system
(\ref{tdp}) follows and
 the expressions involving the product between $v_k$ and
$\dot v_k^*$ apearing in equation  (\ref{lagra}) are determined:
\begin{equation}
\fl T_{k}={i\hbar\over 2}(\dot v_l v_l^*-\dot v_l^*v_l)=2\rho_l(\epsilon_l-\lambda)-
G\rho_l^2+{\Delta_0^*\over 2}\left({\rho_l^2\over\kappa_l^*}-\kappa_l\right)
+{\Delta_0\over 2}\left({\rho_l^2\over\kappa_l}-\kappa_l^*\right).
\label{ttt}
\end{equation}
 The discussion found Refs. \cite{nix} reveal
that the approximation used to obtain Rel. (\ref{eqd}) helps us to describe
approximately the effect of the residual interaction on dissipation and we
eliminate only the terms related to the collective kinetic energy. The collective
kinetic energy will be treated separately by solving equations
involving different seniority states.

In order to determine the excitations between configurations
the equations (\ref{lagra}) must be derived  with respect the amplitudes 
$c_{0}$, $c_{0}^*$, $c_{ij}$, $c_{ij}^{*}$, $c_{jj}$ and
$c_{jj}^*$. The next three equations follows:

\begin{equation}
c_{0}E_{0}+i\hbar \dot c_{0}-c_{0}\sum_{k}T_{k} 
-i\hbar\sum_{l,j\ne l}c_{jl}(u_{j}v_{l}^*-v_{j}^*u_{j})
\langle j\mid{ \partial\over\partial t}\mid l\rangle  = 0;
\label{eqcj1}
\end{equation}

\begin{eqnarray}
c_{jl}E_{jl}-i\hbar \dot{c}_{jl}-c_{jl}\sum_{k\ne j,l}T_{k}
-
i\hbar[c_{0}(-u_{j}v_{l}+v_{j}u_{l})
\langle j\mid {\partial\over\partial t}\mid l\rangle  & 
\label{eqcj}\\
+\sum_{m}c_{ml}(u_{j}u_{m}+v_{j}v_{m}^*)
\langle m\mid {\partial\over\partial t}\mid j\rangle 
+\sum_{n}c_{jn}(u_{n}u_{l}+v_{l}v_{n}^*)
\langle \bar n\mid{ \partial\over\partial t}\mid \bar l\rangle
&=0;\nonumber
\end{eqnarray}

\begin{equation}
c_{jj}(E_{jj}-\sum_{k\ne j}T_{k}+T_{j})-i\hbar\dot c_{jj} 
+i\hbar c_{0}(u_{j}\dot v_{j}-v_j\dot u_j) = 0;
\label{eqcj2}
\end{equation}
and three equations for their complex conjugates.
The notations (\ref{uiy}), (\ref{euri1}), (\ref{euri}), (\ref{tk}) and
(\ref{ttt}) were used.
The homogeneous solutions are:

\begin{eqnarray}
c_{0(h)}(t)&=&c_{0(h)}(0)\exp\left(-{i\over\hbar}\int_{0}^{t}
(E_{0}-\sum_{k}T_{k})dt\right);
\end{eqnarray}

\begin{eqnarray}
c_{jl(h)}(t)&=&c_{jl(h)}(0)\exp\left(-{i\over\hbar}\int_{0}^{t}
(E_{jl}-\sum_{k\ne j,l}T_{k})dt\right);
\end{eqnarray}

\begin{eqnarray}
c_{jj(h)}(t)&=&c_{jj(h)}(0)\exp\left(-{i\over\hbar}\int_{0}^{t}(E_{jj}
-\sum_{k\ne j}T_{k}+T_{j})dt\right).
\end{eqnarray}

If the system deforms slowly, the contributions of the type
$c_{jl}\langle j\mid\partial/\partial t\mid l\rangle$ in
Eqs. (\ref{eqcj1}), (\ref{eqcj}) and (\ref{eqcj2}) can 
be neglected and the solutions 
for seniority-two states given by the Lagrange method 
of variation of constants are:

\begin{equation}
c_{jl}={-i\hbar\over E_{jl}-\sum_{k\ne j,l}T_k-E_{0}+\sum_{k}T_{k}}
(v_ju_l-u_jv_l)\langle j\mid{\partial\over\partial t}\mid l\rangle 
c_{0(h)}
\end{equation}

\begin{eqnarray}
c_{jj}&=&{-i\hbar\over E_{jj}-\sum_{k\ne j}T_k+T_j-E_{0}+\sum_{k}T_{k}}
(u_{j}\dot v_{j}-v_{j}\dot u_{j})c_{0(h)}
\end{eqnarray}

Finally the probabilities to find the system in a seniority-two
state are:

\begin{equation}
\mid c_{jl}\mid^2=\hbar^2{ \mid v_ju_l-u_jv_l\mid^2
\langle j\mid{\partial\over\partial t}\mid l\rangle^2 
\over (E_{jl}-\sum_{k\ne j,l}T_k-E_{0}+\sum_{k}T_{k})^2}
\mid c_{0(h)}\mid^2
\end{equation}
\begin{equation}
\mid c_{jj}\mid^2=\hbar^2{ \mid u_{j}\dot v_{j}-v_{j}\dot u_{j}\mid^2
  \over (E_{jj}-\sum_{k\ne j}T_k+T_j-E_{0}+\sum_{k}T_{k})^2}
\mid c_{0(h)}\mid^2
\end{equation}

The total energy of the system is;
\begin{eqnarray}
E&=&\mid c_{0}\mid^{2}E_{0}+\sum_{j,l}\mid c_{jl}\mid^{2}E_{jl}\nonumber\\
&= & (1-\sum_{j,l}\mid c_{jl}\mid^{2})E_{0}+\sum_{j,l}\mid c_{jl}\mid^{2}E_{jl}\\
& =& E_{0}+\sum_{j,l}\mid c_{jl}\mid^{2}(E_{jl}-E_{0})\nonumber
\end{eqnarray}
If we consider that the collective kinetic energy is
\begin{eqnarray}
\sum_{\nu,\mu}{1\over 2}B_{\nu\mu}\dot q_{\nu}\dot q_{\mu}&=&\sum_{j,l}\mid c_{jl}\mid^{2}(E_{jl}-E_{0}),
\end{eqnarray}
and substituting the time derivative with derivatives with respect the generalized
coordinates 
\begin{equation}
{\partial \over \partial t}=\sum_{\nu}\dot q_{\nu}{\partial \over \partial q_{\nu}}
\end{equation}
and considering that $\mid c_0\mid^2\approx 1$ it follows
\begin{eqnarray}
\fl B_{\nu\mu}=2\hbar^2\sum_{m,n\ne m}{ (E_{mn}-E_0)\mid v_mu_n-u_mv_n\mid^2
\langle m\mid{\partial \over\partial q_{\nu}}\mid n\rangle
\langle n\mid{\partial \over\partial q_{\mu}}\mid m\rangle
 \over (E_{mn}-\sum_{k\ne m,n}T_k-E_{0}+\sum_{k}T_{k})^2}\nonumber\\
 +2\hbar^{2}\sum_{m}
{ (E_{mm}-E_0)( u_{m}{\partial v_{m}\over\partial q_{\nu}}-v_{m}{\partial u_{m}\over\partial q_{\nu}})
( u_{m}{\partial v_{m}^*\over\partial q_{\mu}}-v_{m}^*{\partial u_{m}\over\partial q_{\mu}})
  \over (E_{mm}-\sum_{k\ne m}T_k+T_m-E_{0}+\sum_{k}T_{k})^2}
\label{cr}
\end{eqnarray}
where $B_{\nu\mu}$ are the effective mass parameter.

Using the identities 
\begin{equation}
<i\mid {\partial\over\partial t}\mid j>=
{<i\mid{\partial H\over\partial t}\mid j>
\over \epsilon_{j}-\epsilon_{i}}
\end{equation}
the dependences with respect the derivative of the Hamiltonian are evidenced:
\begin{eqnarray}
\fl B_{\nu\mu}=2\hbar^2\sum_{m,n\ne m}{ (E_{mn}-E_0)\mid v_mu_n-u_mv_n\mid^2
\langle m\mid{\partial H\over\partial q_{\nu}}\mid n\rangle
\langle n\mid{\partial H\over\partial q_{\mu}}\mid m\rangle
 \over (E_{mn}-\sum_{k\ne m,n}T_k-E_{0}+\sum_{k}T_{k})^2(\epsilon_{m}-\epsilon_{n})^2}\nonumber\\
 +2\hbar^{2}\sum_{m}
{ (E_{mm}-E_0)( u_{m}{\partial v_{m}\over\partial q_{\nu}}-v_{m}{\partial u_{m}\over\partial q_{\nu}})
( u_{m}{\partial v_{m}^*\over\partial q_{\mu}}-v_{m}^*{\partial u_{m}\over\partial q_{\mu}})
  \over (E_{mm}-\sum_{k\ne m}T_k+T_m-E_{0}+\sum_{k}T_{k})^2}
\end{eqnarray}
The values of $u_{k}$ and $v_{k}$ are solutions of the 
time dependent pairing equations. 
Using notations (\ref{notatii}), the expression
(\ref{crr}) is eventually obtained.

It must be noticed 
that by replacing  
the energies with the approximate values \cite{ogle}
$E_{jl}\approx\sqrt{(\epsilon_{j}-\lambda)^2+\Delta_{jl}^2}+
\sqrt{(\epsilon_{l}-\lambda)^2+\Delta_{jl}^2}=E_{j}+E_{l}$, 
by neglecting the differences between the sums 
of $T_{k}$-terms and by using the stationary values 
$\tilde v$ and $\tilde u$ in the previous
 expression,
it is straightforward to obtain the usual
cranking mass parameter for an adiabatic BCS state \cite{funny}.
In this context, the following identities are also needed:
\begin{eqnarray}
\tilde u_k^2={1\over 2}(1+{\epsilon_{k}-\lambda\over E_{k}});\nonumber\\
\tilde v_k^2={1\over 2}(1-{\epsilon_{k}-\lambda\over E_{k}});\nonumber\\
\langle i\mid {\partial H\over\partial q}\mid i\rangle=
{\partial \epsilon_i\over\partial q}.
\end{eqnarray}


\begin{thebibliography}{99}
\bibitem{nix1} Nix J R 1972 {\it Ann. Rev. Nucl. Sci.} {\bf 22} 65.
\bibitem{inglis}  Inglis D R 1954 {\it Phys. Rev.} {\bf 96} 1059.
\bibitem{inglis2}Inglis D R 1955 {\it Phys.Rev.} {\bf 97} 701.
\bibitem{schutte}Schutte G and Wilets L 1975 {\it Nucl. Phys.} A {\bf 252} 21.
\bibitem{funny}Brack M,  Damgaard J,  Jensen A,  Pauli H, Strutinsky V and
Wong W 1972 {\it Rev. Mod. Phys.} {\bf 44} 320.
\bibitem{iwamoto1}Iwamoto A and Greiner W 1979 {\it Z. Phys.} A {\bf 292} 301.
\bibitem{iwamoto} Iwamoto A and  Maruhn J A 1979 {\it Z. Phys.} A {\bf 293} 315.
\bibitem{schneider}Schneider V,  Maruhn J A and Greiner W 1986 {\it Z. Phys.} A {\bf 323} 111.
\bibitem{liran} Liran S, Scheefer H J, Scheid W and Greiner W 1975
{\it Nucl. Phys.} A {\bf 248} 191.
\bibitem{chaff} Chaffin E F and Dickmann F 1976 {\it Phys. Rev. Lett}
{\bf 37} 1738.
\bibitem{nix} Koonin S E and  Nix J R 1976 {\it Phys. Rev.} C {\bf 13} 209.
\bibitem{franta}  Blocki J and Flocard H 1976 {\it Nucl. Phys.} A {\bf 273} 45.
\bibitem{mireanp} Mirea M, Tassan-Got L, Stephan C and  Bacri C O 2004 {\it Nucl Phys.} A {\bf 735}
21.
\bibitem{mirealz}Mirea M, Tassan-Got L, Stephan C, Bacri C O and  Bobulescu R C
2007 {\it Phys. Rev.} C {\bf 76} 064608.
\bibitem{mireapl} Mirea M  arXiv:0907.1347v1; {\it Phys. Lett. B} submitted.
\bibitem{mirea2}Mirea M 2008 {\it Phys. Rev.} C {\bf 78} 044618.
\bibitem{davies}Davies K T R and  Nix J R 1976 {\it Phys. Rev.} C {\bf 14}
1977.
\bibitem{ejpa}Mirea M, Bajeat O, Clapier F, Ibrahim F,
 Mueller A C,  Pauwels N and  Proust J 2001 {\it Eur. Phys. J.} A {\bf 11} 
59.
\bibitem{mnpa}  Mirea M 2006 {\it Nucl. Phys.} A {\bf 780} 13.
\bibitem{gr}  Maruhn J and Greiner W 1972 {\it Z. Phys.} {\bf 251} 431.
\bibitem{mirea3}Mirea M, Tassan-Got L, Stephan C, Bacri C O,
Stoica P  and  Bobulescu R C 2005 {\it J. Phys. G} {\bf 31} 1165.
\bibitem{leder}Ledergerber T and Pauli H-C 1973 {\it Nucl Phys} A {\bf 207} 1.
\bibitem{mbm} Mirea M, Bobulescu R C and Marian P {\it Rom. Rep. Phys.} in print.
\bibitem{torres} Diaz-Torres A, Gasques L R and Wiescher M 2007
{\it Phys. Lett} B {\bf 652} 255.
\bibitem{hoff} Hofmann H 1976 {\it Phys. Lett. B} {\bf 61} 423.
\bibitem{brink}Brink D M and Weiguny A 1968 {\it Nucl. Phys. A}
{\bf 120} 59.
\bibitem{gozdz} Gozdz A, Pomorski K, Brack M and Werner E
1985 {\it Nucl Phys. A} {\bf 442} 26.
\bibitem{baran}  Baran A,  Sheikh J A and  Nazarewicz W 2009 
{\it Int. J. Mod. Phys.} E {\bf 18}, 1054.
\bibitem{dudek} Dudek J, Dudek W, Ruchowska E and Skalski J 1980
{\it Z. Phys.} A {\bf 294} 341.
\bibitem{ogle} Ogle W, Wahlborn S, Piepenbring R and 
Fredriksson S 1971 {\it Rev. Mod. Phys} {\bf 42} 424.
\end{thebibliography}
\end{document}